\documentclass[authoryear,final,5p,twocolumn]{elsarticle}

\usepackage{amssymb}
\usepackage{amsmath}
\usepackage{cuted}
\usepackage{array}
\usepackage{graphicx}
\usepackage{caption}
\usepackage{subcaption}
\usepackage{color}

\newtheorem{thm}{Theorem}
\newtheorem{lem}[thm]{Lemma}
\newtheorem{prp}[thm]{Proposition}
\newdefinition{rmk}{Remark}
\newproof{pf}{\textbf{Proof}}

\makeatletter
\def\ps@pprintTitle{%
   \let\@oddhead\@empty
   \let\@evenhead\@empty
   \let\@oddfoot\@empty
   \let\@evenfoot\@oddfoot
}
\makeatother

\begin{document}

\begin{frontmatter}

\title{Handling plant-model mismatch in Koopman Lyapunov-based model predictive control via offset-free control framework}
\author[a,b]{Sang Hwan Son}\ead{zshson@gmail.com}
\author[a,b]{Abhinav Narasingam}\ead{abhinavn0708@tamu.edu}
\author[a,b]{Joseph Sang-Il Kwon\corref{cor}}\ead{kwonx075@tamu.edu}
\address[a]{Artie McFerrin Department of Chemical Engineering, Texas A\&M University, College Station, TX 77845 USA}
\address[b]{Texas A\&M Energy Institute, Texas A\&M University, College Station, TX 77845 USA}
\cortext[cor]{Corresponding author J.~S.-I.~Kwon. Tel. +1-979-862-5930. 
Fax +1-979-845-6446.}

\begin{abstract}
Koopman operator theory enables a global linear representation of a given nonlinear dynamical system by transforming the nonlinear dynamics into a higher dimensional observable function space where the evolution of observable functions is governed by an infinite-dimensional linear operator. For practical application of Koopman operator theory, various data-driven methods have been developed to derive lifted state-space models via approximation to the Koopman operator. Based on approximate models, several Koopman-based model predictive control (KMPC) schemes have been proposed. However, since a finite-dimensional approximation to the infinite-dimensional Koopman operator cannot fully represent a nonlinear dynamical system, plant-model mismatch inherently exists in these KMPC schemes and negatively influences the performance of control systems. In this work, we present offset-free Koopman Lyapunov-based model predictive control (KLMPC) framework that addresses the inherent plant-model mismatch in KMPC schemes using an offset-free control framework based on a disturbance estimator approach and ensures feasibility and stability of the control system by applying Lyapunov constraints to the optimal control problem. The zero steady-state offset condition of the developed framework is mathematically examined. The effectiveness of the developed framework is also demonstrated by comparing the closed-loop results of the proposed offset-free KLMPC and the nominal KLMPC.
\end{abstract}


\end{frontmatter}
	
\section*{Introduction} \label{intro}

Koopman and Von Neumann showed that a nonlinear dynamical system can be represented by an operator-theoretic perspective, which is to describe the system in terms of functions of states (called observables). They also showed that the temporal evolution of these observables can be described by an infinite-dimensional linear operator called Koopman operator \citep*{17,18}.

Recently, based on the availability of abundant time-series data and advances in numerical techniques, several data-driven methods have been developed to obtain a finite-dimensional approximation to the infinite-dimensional Koopman operator for practical application of Koopman operator theory \citep*{24,29,38}. Dynamic mode decomposition (DMD) proposed in \cite*{21,26} produces an approximation of the Koopman operator using a set of monomials. DMD has been applied in many fields by the benefit of its ease of implementation. However, since DMD only utilizes a limited space that consists of scalar observables, DMD cannot provide proper prediction power for highly nonlinear systems. Extended dynamic mode decomposition (EDMD) is developed in \cite*{21,26} to handle this limitation of DMD. EDMD enriches the observable space by using nonlinear functions and provides a proper numerical approximation of Koopman operator. 

A key feature of data-driven model identification methods based on Koopman operator theory is the availability of linear predictors in lifted space for nonlinear dynamical systems. These linear predictors allow for implementation of well-established linear controller design methods to nonlinear systems such as linear model predictive control (MPC) that has been applied in various applications as a standard model-based control strategy \citep*{32,30,31}. Based on this feature, various Koopman-based model predictive control (KMPC) schemes have been developed and applied for control of several nonlinear systems \citep*{34,33,37}. \cite*{11} proposed Koopman Lyapunov-based model predictive control (KLMPC) that integrates KMPC with Lyapunov-based MPC (LMPC) scheme presented in \cite*{27,28} to guarantee feasibility and stability of control systems. Additionally, \cite*{16} presented a mathematical analysis for feasibility and stability of KLMPC system in the Lyapunov sense.

Since a finite-dimensional approximation obtained from previously introduced data-driven numerical methods (e.g.,~EMDM) cannot completely represent the infinite-dimensional Koopman operator, plant-model mismatch is inevitable in KLMPC. This inherent plant-model mismatch in KLMPC can significantly degrade the closed-loop performance of the KLMPC systems, but to the best of our knowledge, handling of this plant-model mismatch in KLMPC system has not been rigorously studied yet. To this end, we propose an offset-free KLMPC framework that integrates KLMPC with the offset-free MPC framework to address the inherent plant-model mismatch in KLMPC while guaranteeing feasibility and stability of a control system with Lyapunov constraints.

Offset-free MPC is one of the most representative plant-model mismatch compensating schemes in the MPC field. There are two ways to achieve offset-free tracking in the presence of plant-model mismatch. One method exploits the integration of tracking error in a compensator block as in \cite*{35,36}. However, since the integrated error is independent of the controller, this method can cause a windup problem where an overshoot occurs due to an accumulated error even after the system reaches a desired point. Second approach is to design a disturbance estimator to compensate for plant-model mismatch by augmenting a disturbance model to the system model and estimating disturbance-augmented state from measurement \citep*{2,4}. Since the disturbance estimator approach has an anti-windup effect \citep*{6}, this approach is one of the most popular approaches to accomplish offset-free tracking in MPC \citep*{1}.

Therefore, in this study, we develop an offset-free KLMPC framework based on the disturbance estimator approach. Specifically, a disturbance model is introduced to consider the influence of plant-model mismatch, and is augmented with a linear model in the lifted state space, identified by EDMD. Then, based on the disturbance-augmented model, an estimator is designed to estimate the lifted state and disturbance. A target problem is also designed to obtain proper target values for the lifted states and inputs that will be applied to the optimal control problem by considering the estimated disturbance. Specifically, the optimal control problem is designed based on the disturbance-augmented model and Lyapunov constraints to obtain an optimal input to track the set-point while compensating for the influence of plant-model mismatch on the controlled variables and ensuring the feasibility and stability of control system. Unlike the nominal KLMPC scheme, since the target lifted state and input values of the optimal control problem are continuously updated through the target problem, the Lyapunov function and stabilizing control law within the Lyapunov constraints are also continuously updated during the closed-loop operation. A mathematical analysis for zero steady-state offset condition of the proposed offset-free KLMPC framework is also presented.

The rest of this paper is organized as follows. The next section introduces a data-driven EDMD scheme to derive linear models in the lifted state-space for nonlinear dynamical systems and presents a formulation of the nominal KLMPC scheme that integrates EDMD and Lyapunov-based control. In Section~\ref{sec2}, a detailed formulation of the offset-free KLMPC framework and mathematical analysis for zero stead-state offset condition are presented. In Section~\ref{sec3}, the performance of the proposed framework in handling plant-model mismatch compared to the nominal KLMPC is shown with a closed-loop simulation of a CSTR system. In the last section, we conclude with a few important remarks.

\section{Preliminaries} \label{sec1}

\subsection{Data-driven linear model identification of nonlinear systems} \label{sec1_1}

In this section, we introduce a data-driven model identification method to obtain a linear model in a higher dimensional lifted space of nonlinear systems, based on Koopman operator theory.

In \cite*{17,18}, Koopman suggested an alternative operator theoretic perspective to describe the dynamics of an uncontrolled system in (\ref{eq1}) in terms of the evolution of observable functions $\psi:\mathbb{R}^{n_x}\rightarrow \mathbb{R}$.
\begin{align}\label{eq1}
x(k+1)=f(x(k))
\end{align}
where $x\in\mathbb{R}^{n_{x}}$ represents the state, and $n_x$ denotes the dimension of state. 

Koopman has also shown that an infinite-dimensional linear operator $\mathcal{K}:\mathcal{F}\rightarrow \mathcal{F}$ exists which advances the observables forward in time as follows:
\begin{align}\label{eq2}
\mathcal{K}\psi(x)=\psi(f(x)).
\end{align}
where $\mathcal{F}$ is a space of observables invariant under the action of the Koopman operator \citep*{20}.

Since the Koopman operator theory was initially introduced based on uncontrolled systems, several schemes have been suggested to generalize the Koopman operator theory to controlled systems of the following form \citep*{26,22,19}:
\begin{align}\label{eq3}
x(k+1)=f(x(k),u(k))
\end{align}
where $u\in\mathcal{U}\subset\mathbb{R}^{n_u}$ represents the input, and $n_u$ denotes the dimension of input. 

In this study, we present a scheme in \cite*{19} that generalizes the Koopman operator theory for controlled systems by introducing an extended state as
\begin{align}\label{eq4}
\chi=\begin{bmatrix} x \\ \mathbf{u} \end{bmatrix}
\end{align}
where $\mathbf{u}:=\{ u_i \}^\infty_{i=0}\in\ell(\mathcal{U})$ ($u_i\in\mathcal{U}$) represents the input sequence and $\ell(\mathcal{U})$ denotes the space of all input sequences $\mathbf{u}$. The dynamics of the extended state $\chi$ is described as
\begin{align}\label{eq5}
f_\chi(\chi)=\begin{bmatrix} f(x,\mathbf{u}(0)) \\ \mathcal{S}\mathbf{u} \end{bmatrix}
\end{align}
where $\mathbf{u}(i)$ denotes the $i^{th}$ element of $\mathbf{u}$ and $\mathcal{S}$ represents the left shift operator:
\begin{align}\label{eq6}
(\mathcal{S}\mathbf{u})(i)=\mathbf{u}(i+1).
\end{align}

Then, the Koopman operator $\mathcal{K}:\mathcal{H}\rightarrow\mathcal{H}$ associated with the dynamics of the extended state in (\ref{eq5}) can be defined based on an extended observable $\phi:\mathbb{R}^{n_x}\times\ell(\mathcal{U})\rightarrow\mathbb{R}$:
\begin{align}\label{eq7}
\mathcal{K}\phi(\chi)=\phi(f_\chi(\chi))
\end{align}
where $\mathcal{H}$ is an extended observable space.

Since the Koopman operator itself is not suitable for practical implementation due to its infinite dimensional nature, several schemes have been developed for finite-dimensional approximation to the Koopman operator. In this work, we present a data-driven EDMD algorithm \citep*{21,26}.

In EDMD, a vector of observables $\phi$ is designed as
\begin{align}
&\phi(x,\mathbf{u})=\begin{bmatrix} \psi(x) \\ \mathbf{u}(0) \end{bmatrix}\label{eq8}\\
&\psi(x):=\begin{bmatrix} \psi_1(x),\cdots,\psi_{n_z}(x) \end{bmatrix}^\top\nonumber
\end{align}
where $n_z$ is the number of observables depending on $x$.

The scheme assumes that a collection of data $((x_j,\mathbf{u}_j),(x_j^+,\mathbf{u}_j^+))$, $j=1,\cdots,N_d$ that satisfying (\ref{eq9}) is available.
\begin{align}\label{eq9}
\begin{bmatrix} x_j^+ \\ \mathbf{u}_j^+ \end{bmatrix}= \begin{bmatrix} f(x_j,\mathbf{u}_j(0)) \\ \mathcal{S}\mathbf{u}_j \end{bmatrix}
\end{align}
where the superscript $+$ denotes the value at the next time step. Then, based on the data, an approximation of the Koopman operator $\mathcal{A}$ is obtained by minimizing
\begin{align}\label{eq10}
J(\mathcal{A})=\sum_{j=1}^{N_d} || \phi(x_j^+,\mathbf{u}_j^+)- \mathcal{A}\phi(x_j,\mathbf{u}_j)||^2
\end{align}
where $||\cdot||$ denotes the Euclidean norm. 

Since it is not essential to predict the future control sequence, we can discard the last $n_u$ components of each $\phi(x_j^+,\mathbf{u}_j^+)$ and the last $n_u$ rows of $\mathcal{A}$. Additionally, let $\bar{\mathcal{A}}$ denote the remaining part of $\mathcal{A}$ after discarding the part associated with the future input, then we can decompose $\bar{\mathcal{A}}$ into $A\in\mathbb{R}^{n_z \times n_z}$ and $B\in\mathbb{R}^{n_z \times n_u}$ as
\begin{align}\label{eq11}
\bar{\mathcal{A}}=\begin{bmatrix} A & B \end{bmatrix}.
\end{align}
Then, substituting (\ref{eq11}) into the remaining part of (\ref{eq10}) yields
\begin{align}\label{eq12}
\bar{J}(A,B)=\sum_{j=1}^{N_d} || \psi(x_j^+)- A\psi(x_j)-B\textbf{u}_j(0) ||^2.
\end{align}
Finally, we can obtain a linear predictor in lifted space as in (\ref{eq13}) by minimizing (\ref{eq12}) over $A$ and $B$.
\begin{align}\label{eq13}
z(k+1)=A z(k)+B u(k)
\end{align}
where $z\in\mathbb{R}^{n_z}$ denotes the lifted state:
\begin{align}\label{eq14}
z=\psi(x).
\end{align}

Additionally, let the output be given as a function of the state as
\begin{align}\label{eq15}
y(k)=g(x(k)).
\end{align}
In this case, the output matrix $C$ is obtained as the best projection of $y$ onto the span of observable functions in a least square sense by minimizing
\begin{align}\label{eq16}
J_y(C)=\sum_{j=1}^{N_d} || y_j-C\psi(x_j)||^2.
\end{align}
Then, a linear model for $y$ can be derived using $C$ as follows:
\begin{align}\label{eq17}
y(k)=Cz(k).
\end{align}

\subsection{KLMPC design}\label{sec1_2}

Based on the data-driven model identification methodology presented in the previous section, \cite*{11} developed KLMPC framework that integrates LMPC scheme in \cite*{27,28} with EDMD to ensure feasibility and stability of control systems.

In LMPC, following Lyapunov constraints are applied to the optimal control problem:
\begin{align}
&V(x_{i+1}-\bar{x}_s)\leq r,\quad i=0,\dots ,N-1\label{eq18}\\
&V(x_{1}-\bar{x}_s)\leq V(x_1^h-\bar{x}_s)\label{eq19}
\end{align}
where $V(x-\bar{x}_s)$ represents the Lyapunov function, $\bar{x}_s$ denotes the target steady-state point of the control system, $x_1,\cdots,x_{N-1}$ denote future states, and $x_1^h$ denotes the state evolved from $x_0$ with a stabilizing control law. The first Lyapunov constraint in (\ref{eq18}) guarantees that future states stay within the stability region which can be described as a sublevel set of the Lyapunov function $\Omega_r:=\{ x\in{\mathbb{R}^{n_x}}:V(x-\bar{x}_s)\leq r \}$. Then, the second Lyapunov constraint in (\ref{eq19}) ensures that the Lyapunov function value at the next time step induced by the MPC control law is smaller than or equal to that induced by a stabilizing control law. These Lyapunov constraints ensure the closed-loop stability of LMPC systems.

However, since these Lyapunov constraints are nonlinear, the optimal control problem becomes a non-convex problem. \cite*{11,16} effectively address this issue by transforming the nonlinear Lyapunov constraints in (\ref{eq18}) and (\ref{eq19}) to linear constraints with EDMD. Specifically, the Lyapunov function is incorporated into the observable function library and described as one of the lifted states. 

For example, when the Lyapunov function is a quadratic form, it can be incorporated into the observable function library as the $j_v^{th}$ observable:
\begin{align}\label{eq20}
\psi_{j_v}(x)=(x-\bar{x}_s)^\top Q_v(x-\bar{x}_s).
\end{align}
Then, after obtaining the linear model in the lifted space in (\ref{eq13}) through EDMD, the Lyapunov function value can be simply derived from the lifted state as follows:
\begin{align}\label{eq21}
V(x-\bar{x}_s)=D_{v} z
\end{align}
where $D_v:=\begin{bmatrix} \mathbf{0}_{1,j_v-1}, 1,\mathbf{0}_{1,n_z-j_v} \end{bmatrix}$. The Lyapunov constraints can be reformulated as (\ref{eq22}) and (\ref{eq23}) by substituting (\ref{eq21}) into (\ref{eq18}) and (\ref{eq19}).
\begin{align}
&D_{v} z_{i+1} \leq r \label{eq22}\\
&D_{v} z_1 \leq V(D_{x} (z_1^h-\bar{z}_s)) \label{eq23}
\end{align}
where 
\begin{align}
&D_{x}:=\begin{bmatrix} D_{{x_1}}^\top,\cdots,D_{{x_{n_x}}}^\top \end{bmatrix}^\top\nonumber\\
&D_{x_j}:=\begin{bmatrix} \mathbf{0}_{1,j-1}, 1,\mathbf{0}_{1,n_z-j} \end{bmatrix}\quad \mathrm{for} \quad j=1,\cdots,n_x\nonumber\\
&\bar{z}_s:=\psi(\bar{x}_s). \nonumber
\end{align}

Then, the finite-horizon optimal control problem of KLMPC can be constructed as a standard convex quadratic problem in (\ref{eq24}) by applying the Lyapunov constraints in (\ref{eq22}) and (\ref{eq23}):
\begin{subequations}\label{eq24}
\begin{align}
\underset{\small{u_0,\cdots,u_{N-1}}}{\mathrm{min}}&\; \sum _{ i=0 }^{ N-1 }{||z_{i+1}-\bar{z}_s||^2_{Q_z}+||u_i-\bar{u}_s||^2_{Q_{u}}} \label{eq24a}\\
\mathrm{s.t.}\quad 
&z_{0}=z\label{eq24b}\\
&z_{i+1}=Az_i +Bu_i\label{eq24c}\\
&z_1^h=Az_0+Bh_0(z_0,\bar{z}_s)\label{eq24d}\\
&u_{min} \leq u_i \leq u_{max} \label{eq24e}\\
&y_{min} \leq Cz_{i+1} \leq y_{max} \label{eq24f}\\
&D_{v} z_{i+1} \leq r,\quad i=0,\dots ,N-1\label{eq24g}\\
&D_{v} z_1 \leq V(D_{x} (z_1^h-\bar{z}_s))\label{eq24h}
\end{align}
\end{subequations}
where $\bar{u}_s$ is the steady-state target input.

\section{Offset-free Koopman Lyapunov-based MPC}\label{sec2}

The KLMPC scheme in the previous section provides controller feasibility and stability by effectively integrating EDMD and LMPC. However, the closed-loop performance of a KLMPC system can be degraded due to the plant-model mismatch between a real plant dynamics in (\ref{eq3}) and a data-driven linear model, obtained by EDMD, in (\ref{eq13}). To this end, we present an offset-free KLMPC framework that integrates an offset-free MPC framework in \cite*{6,7} with the KLMPC scheme to compensate for this plant-model mismatch.

\subsection{Offset-free KLMPC framework}\label{sec2_1}

We suppose a data-driven linear model for lifted state $z$ in (\ref{eq13}) and (\ref{eq17}) has been identified by EDMD in Section~\ref{sec1_1} and the controlled variable is described as
\begin{align}\label{eq25}
y_c(k)=Hy(k)
\end{align}
where $y_c\in\mathbb{R}^{n_{y_c}}$ is the controlled variable and $n_{y_c}$ denote the dimension of the controlled variable.

In the disturbance estimator approach, a disturbance model is augmented with the lifted state-space model to consider the effect of plant-model mismatch. Though various disturbance-augmented models have been proposed in \cite*{12,13,14}, in this study, we focus on the following form \citep*{4}:
\begin{align}\label{eq26}
\begin{cases} z(k+1)=A z(k)+B u(k)+B_d d(k) \\ d(k+1)=d(k) \\ y(k)=C z (k)+C_d d(k) \end{cases}
\end{align}
where $d\in\mathbb{R}^{n_d}$ is the disturbance, and $B_d\in\mathbb{R}^{n_z\times n_d}$ and $C_d\in\mathbb{R}^{n_y\times n_d}$ denote matrices that represent the influence of the disturbance variable on the evolution of the lifted state and the output, respectively.

$B_d$ and $C_d$ can be freely determined under the restriction that the disturbance-augmented system in (\ref{eq26}) is observable. The observability condition of the augmented system associated with $B_d$ and $C_d$ design is given in the following proposition.

\begin{prp}\label{prp1}
The disturbance-augmented system in (\ref{eq26}) is observable if and only if the original linear dynamical system in (\ref{eq13}) is observable and the following condition holds. The detailed derivation of the condition is provided in Proposition 1 of \cite*{1}.
\begin{align}\label{eq27}
{\mathrm{rank}} \begin{bmatrix} I-A & -B_d \\ C & C_d \end{bmatrix} =n_z+n_d.
\end{align}
\end{prp}

\begin{rmk} \label{rmk1}
\textit{For the condition in (\ref{eq27}) to be satisfied, the dimension of the disturbance should be smaller than or equal to the number of available measurements \citep*{5,15}:
\begin{align}\label{eq28}
n_d \leq n_y.
\end{align}}
\end{rmk}

The disturbance-augmented model with properly designed $B_d$ and $C_d$ allows a model predictive controller to effectively handle the negative impact of plant-model mismatch on the lifted state and the output. The following lifted state and disturbance estimator is designed based on the constructed model in (\ref{eq26}).
\begin{align}\label{eq29}
\begin{bmatrix} \hat{z}(k+1) \\ \hat{d}(k+1) \end{bmatrix}&=
\begin{bmatrix} A & B_d \\ 0 & I \end{bmatrix} \begin{bmatrix} \hat{z}(k) \\ \hat{d}(k) \end{bmatrix}+ \begin{bmatrix} B \\ 0 \end{bmatrix} u(k) \nonumber\\
& + \begin{bmatrix} L_z \\ L_d \end{bmatrix} (-y_p(k)+C\hat{z}(k)+C_d\hat{d}(k))
\end{align}
where $\hat{z}$ and $\hat{d}$ represent the estimated lifted state and disturbance, respectively, $L_z$ and $L_d$ are the estimator gains for lifted state and disturbance, respectively, that make the estimator stable, and $y_p$ denotes the measured output from the plant.

Then, the equilibrium target lifted state $\bar{z}$ and target input $\bar{u}$ that satisfy the following equations are obtained:
\begin{align} 
\bar{z}=A\bar{z}+B\bar{u}+B_d\hat{d} \label{eq30}\\
\bar{y}_c=HC\bar{z}+HC_d\hat{d}.\label{eq31}
\end{align}
We can construct the following equation from (\ref{eq30}) and (\ref{eq31}):
\begin{align}\label{eq32}
\begin{bmatrix} A-I & B \\ HC & 0 \end{bmatrix} \begin{bmatrix} \bar{z} \\ \bar{u} \end{bmatrix} = \begin{bmatrix} -B_d \hat{d} \\ \bar{y}_c -HC_d \end{bmatrix}.
\end{align}

A solution of (\ref{eq32}) exists with proper $H$ and $\bar{y}_c$, but it may not be unique. Therefore, a set of target lifted state and target input $(\bar{z},\bar{u})$ is obtained by solving the following target problem based on desired target values $(\bar{z}_s,\bar{u}_s)$ and process constraints:
\begin{subequations}\label{eq33}
\begin{align}
\underset{\small{\bar{z},\bar{u}}}{\mathrm{min}}\quad &||\bar{u}-\bar{u}_{s}||^2_{Q_{\bar{u}}}+||\bar{z}-\bar{z}_{s}||^2_{Q_{\bar{z}}} \label{eq33a}\\
\mathrm{s.t.}\quad &\begin{bmatrix} A-I & B \\ HC & 0 \end{bmatrix}
\begin{bmatrix} \bar{z} \\ \bar{u} \end{bmatrix}
= \begin{bmatrix} -B_d \hat{d} \\ \bar{y}_c-HC_d \hat{d} \end{bmatrix}\label{eq33b}\\
&u_{min} \leq \bar{u} \leq u_{max} \label{eq33c}\\
&y_{min} \leq \bar{y} \leq y_{max}.\label{eq33d}
\end{align}
\end{subequations}
where $\bar{y}=C\bar{z}+C_d \hat{d}$.

\begin{rmk}\label{rmk2}
\textit{In the nominal offset-free MPC framework, the objective function of the target problem in (\ref{eq33a}) is commonly designed to drive the target output value $\bar{y}$ to a desired value $\bar{y}_{s}$ \citep*{2,3,5}. For example, the following objective function is commonly utilized in nominal offset-free MPC.
\begin{align}\label{eq34}
&\bar{J}_0=||\bar{u}-\bar{u}_s||^{2}_{Q_{\bar{u}}}+||\bar{y}-\bar{y}_{s}||^{2}_{Q_{\bar{y}}}
\end{align}
This objective function induces the state target $\bar{x}$ to stay near a desired state $\bar{x}_s$ that satisfies
\begin{align}\label{eq35}
C_x \bar{x}_s+C_d\hat{d}=\bar{y}_{s}.
\end{align}
However, in the case of the proposed offset-free KLMPC framework, since the system for the lifted state in (\ref{eq26}) from EDMD is higher dimensional than the original system in (\ref{eq3}), the influence of the output correction on the lifted state target $\bar{z}$ is insignificant. Therefore, we replaced the output correction term $||\bar{y}-\bar{y}_{s}||^{2}_{Q_{\bar{u}}}$ in the objective function in (\ref{eq34}) with the lifted state correction term $||\bar{z}-\bar{z}_{s}||^2_{Q_{\bar{z}}}$ as in (\ref{eq33a}) to directly drive the target lifted state to a desired point. The derivation of a proper target lifted state $\bar{z}$ with this setting is crucial because the derived target value will be applied to a finite-horizon optimal control problem.}
\end{rmk}
 
As described in Section~\ref{sec1_2}, the Lyapunov constraints in (\ref{eq18}) and (\ref{eq19}) are easily handled in the KLMPC scheme by including the Lyapunov function with fixed target $\bar{x}_s$ as a component of the observable function library. However, in the proposed offset-free KLMPC framework, since the target lifted state $\bar{z}$ of the optimal control problem is continuously updated over time through the target problem in (\ref{eq33}) as the estimated disturbance $\hat{d}$ is continuously updated from the estimator in (\ref{eq29}), the following Lyapunov constraints also should be continuously updated.
\begin{align}
&V(x_{i+1}-\bar{x})\leq r,\quad i=0,\dots ,N-1\label{eq36}\\
&V(x_{1}-\bar{x})\leq V(x_1^h-\bar{x})\label{eq37}\\
&\bar{x}:=D_x \bar{z} \nonumber\\
&x_1^h:=D_x z_1^h \nonumber
\end{align}
where $z_1^h$ is the lifted state evolved from $z_0$ by the stabilizing control law $h(z_0,\hat{d},\bar{z})$:
\begin{align}\label{eq38}
z_1^h=Az_0 +Bh(z_0,\hat{d},\bar{z})+B_d \hat{d}.
\end{align}

Since the equilibrium point $\bar{x}$ in the Lyapunov function $V(x-\bar{x})$ is continuously updated, we cannot simply replace $V(x-\bar{x})$ with the observable $\psi_{j_v}=(x-\bar{x}_s)^\top Q_v(x-\bar{x}_s)$ in (\ref{eq20}) which is the Lyapunov function with the fixed equilibrium point $\bar{x}_s$ as in the nominal KLMPC. However, we can describe $V(x-\bar{x})$ using $\psi_{j_v}$ with other terms associated with the update of $\bar{x}$. More specifically, suppose the Lyapynov function is in the following quadratic form:
\begin{align}\label{eq39}
V(x-\bar{x})=(x-\bar{x})^\top Q_v(x-\bar{x}).
\end{align}
Then, $V(x-\bar{x})$ can be reformulated with $\psi_{j_v}$ as follows:
\begin{align}\label{eq40}
V(x-\bar{x})= \psi_{j_v}+2(\bar{x}_s-\bar{x})^\top Q_v x+ \bar{x}^\top Q_v \bar{x}-\bar{x}_s^\top Q_v \bar{x}_s.
\end{align}
Substituting $x=D_x z$, $\bar{x}_s=D_x \bar{z}_s$, and $\psi_{j_v}=D_v z$ into (\ref{eq40}) yields
\begin{align}\label{eq41}
&V(x-\bar{x})=F_v z+\bar{z}^\top D_x^\top Q_v D_x\bar{z}-\bar{z}_s^\top D_x^\top Q_v D_x\bar{z}_s \\
&F_v:=D_{v}+2(\bar{z}_s-\bar{z})^\top D_{x}^\top Q_v D_{x} \nonumber.
\end{align}
With (\ref{eq41}), we can effectively update the Lyapunov function as the target lifted state $\bar{z}$ is updated. Additionally, by applying (\ref{eq41}) to (\ref{eq36}) and (\ref{eq37}), the Lyapunov constraints can be described as linear inequalities as follows:
\begin{align}
&F_v z_{i+1} \leq r-\bar{z}^\top D_x^\top Q_v D_x \bar{z} +\bar{z}_s^\top D_x^\top Q_v D_x \bar{z}_s \label{eq42}\\
&F_v z_1 \leq V(D_x (z_1^h-\bar{z}))-\bar{z}^\top D_x^\top Q_v D_x\bar{z} +\bar{z}_s^\top D_x^\top Q_v D_x \bar{z}_s.\label{eq43}
\end{align}

Not only the Lyapunov function, but also the stabilizing control law $h(z,\hat{d},\bar{z})$ in (\ref{eq38}) should be continuously updated over time as the target lifted state $\bar{z}$ and estimated disturbance $\hat{d}$ are updated. Therefore, $h(z,\hat{d},\bar{z})$ is updated based on the following state feedback law \citep*{10} as follows:
\begin{align}\label{eq44}
h(z,d,\bar{z})=\bar{N}\bar{z}-K_z z-K_d d
\end{align}
where $\bar{N}$ and $K_d$ are gains associated with $\bar{z}$ and $d$, respectively, and $K_z$ is the feedback gain for the lifted state that makes the controlled system stable.

\begin{prp} \label{prp2}
The stabilizing control law in (\ref{eq44}) yields perfect set-point tracking at steady state with the following $\bar{N}$ and $K_d$.
\begin{align}
&\bar{N}=[(I-A+BK_z)^{-1}B]^{\dagger} \label{eq45}\\
&K_d=\bar{N}(I-A+BK_z)^{-1}B_d\label{eq46}
\end{align}
where $\dagger$ denotes the pseudo inverse.
\end{prp}

\begin{pf} Substituting $h(z,d,\bar{z})$ in (\ref{eq44}) into the linear dynamical system in (\ref{eq26}) yields the following controlled system:
\begin{align}
z(k+1)=(A-BK_z)z(k)+B\bar{N}\bar{z}(k)+(B_d-BK_d)d(k).\label{eq47}
\end{align}

Let $z_\infty^h$, $\bar{z}_\infty$, and $d_\infty$ represent the steady-state lifted state, target lifted state, and disturbance of the controlled system in (\ref{eq47}). At steady state, applying $z_\infty^h$, $\bar{z}_\infty$, and $d_\infty$ to (\ref{eq47}) and rearranging yields
\begin{align}\label{eq48}
z_\infty^h=(&I-A+BK_z)^{-1}B\bar{N}\bar{z}_\infty\\
&+(I-A+BK_z)^{-1}(B_d-BK_d)d_\infty \nonumber.
\end{align}
Then, substituting $\bar{N}$ and $K_d$ in (\ref{eq45}) and (\ref{eq46}) into (\ref{eq48}) yields
\begin{align}\label{eq49}
z_\infty^h=\bar{z}_\infty.
\end{align}
Therefore, we can see the controlled system with the stabilizing control law yields perfect set-point tracking at steady state.\qed
\end{pf}

Given the targets ($\bar{z}$, $\bar{u}$) from the target problem in (\ref{eq33}) and estimates ($\hat{z}$, $\hat{d}$) from the estimator in (\ref{eq29}), a finite-horizon optimal control problem for the offset-free KLMPC can be brought into a lifted state-regulation problem $P(\hat{z},\hat{d},\bar{z},\bar{u})$ as follows:
\begin{subequations}\label{eq50}
\begin{align}
\underset{\small{u_0,\cdots,u_{N-1}}}{\mathrm{min}}&\; \sum _{ i=0 }^{ N-1 }{||z_{i+1}-\bar{z}||^2_{Q_z}+||u_i-\bar{u}||^2_{Q_u}} \label{eq50a}\\
\mathrm{s.t.}\quad 
&z_{0}=\hat{z}\label{eq50b}\\
&z_{i+1}=Az_i +Bu_i+B_d \hat{d}\label{eq50c}\\
&z_1^h=Az_0+Bh(z_0,\hat{d},\bar{z})+B_d \hat{d}\label{eq50d}\\
&u_{min} \leq u_i \leq u_{max} \label{eq50e}\\
&y_{min} \leq Cz_{i+1}+C_d \hat{d} \leq y_{max} \label{eq50f}\\
&F_v z_{i+1} \leq r-\bar{z}^\top D_x^\top Q_v D_x \bar{z} \label{eq50g}\\ 
 &\qquad\quad+\bar{z}_s^\top D_x^\top Q_v D_x \bar{z}_s,\quad i=0,\dots ,N-1 \nonumber\\
&F_v z_1 \leq V(D_x (z_1^h-\bar{z}))-\bar{z}^\top D_x^\top Q_v D_x\bar{z} \label{eq50h} \\
 &\qquad\quad+\bar{z}_s^\top D_x^\top Q_v D_x \bar{z}_s. \nonumber
\end{align}
\end{subequations}
The Lyapunov constraints in (\ref{eq50g}) and (\ref{eq50h}) are formulated based on (\ref{eq42}), (\ref{eq43}), and (\ref{eq44}). By solving the optimal control problem, we obtain an optimal input to move the original system in (\ref{eq3}) to steady state where the zero steady-state offset is satisfied in the presence of plant-model mismatch.

\subsection{Zero steady-state offset condition}\label{sec2_2}
In this section, we derive the condition for zero steady-state offset of the proposed offset-free KLMPC by examining the closed-loop behavior of the control system at steady state ($\hat{z}(k)\rightarrow \hat{z}_\infty$, $\hat{d}(k)\rightarrow \hat{d}_\infty$, $\bar{z}(k)\rightarrow \bar{z}_\infty$, $\bar{u}(k)\rightarrow \bar{u}_\infty$, $u(k)\rightarrow u_\infty$, and $y_p(k)\rightarrow y_{p,\infty}$ as $k\rightarrow \infty$).

By considering the estimator in (\ref{eq29}) at steady state, (\ref{eq51}) and (\ref{eq52}) are obtained.
\begin{align}
&\hat{z}_\infty=A\hat{z}_\infty+Bu_\infty+B_d\hat{d}_\infty-L_ze_{y,\infty}\label{eq51}\\
&0=L_de_{y,\infty}\label{eq52}
\end{align}
where $u_\infty$ is an optimal control law derived from the optimal control problem in (\ref{eq50}), and $e_{y,\infty}$ is the error between predicted output and measured output:
\begin{align}
&e_{y,\infty}=y_{p,\infty}-\hat{y}_\infty\label{eq53}
\end{align}
where $\hat{y}_\infty:=C\hat{z}_{\infty}+C_d\hat{d}_{\infty}$. Similarly, (\ref{eq54}) and (\ref{eq55}) are derived by considering (\ref{eq30}) and (\ref{eq31}) at steady state as follows: 
\begin{align}
&\bar{z}_\infty=A\bar{z}_\infty+B\bar{u}_\infty+B_d\hat{d}_\infty\label{eq54}\\
&\bar{y}_c=HC\bar{z}_\infty+HC_d\hat{d}_\infty.\label{eq55}
\end{align}

Now, let $P(\hat{z}_\infty,\hat{d}_\infty,\bar{z}_\infty,\bar{u}_\infty)$ denote the optimal control problem in (\ref{eq50}) at steady state with the converged lifted state estimate $\hat{z}_\infty$, disturbance estimate $\hat{d}_\infty$, lifted state target $\bar{z}_\infty$, and input target $\bar{u}_\infty$. Then, we derive the relation between the optimal control input $u_\infty$ from $P(\hat{z}_\infty,\hat{d}_\infty,\bar{z}_\infty,\bar{u}_\infty)$ and the lifted state estimate $\hat{z}_\infty$ in the following lemma.

\begin{lem}\label{lem3}
Let $P_{\hat{z}_\infty}$ represent the optimal control problem $P(\hat{z}_\infty,\hat{d}_\infty,\bar{z}_\infty,\bar{u}_\infty)$. Without loss of generality, we assume that the set-point is far from the upper and lower bounds of process constraints, and thus, the process constraints in (\ref{eq50e}) and (\ref{eq50f}) and the first Lyapunov constraint in (\ref{eq50g}) are inactive at the optimum of $P_{\hat{z}_\infty}$. Then, the optimal control law of $P(\hat{z}_\infty,\hat{d}_\infty,\bar{z}_\infty,\bar{u}_\infty)$ can be described as
\begin{align}
u_\infty = K_{mpc}\hat{z}_\infty +c_{mpc}\label{eq56}
\end{align}
with
\begin{align}
(K_{mpc},c_{mpc}) = \begin{cases}
(K_{un},c_{un}) &\mathrm{when \; (\ref{eq50h})\; is\; inactive}\\
(K_{\ell},c_{\ell}) &\mathrm{when \; (\ref{eq50h})\; is\; active} \end{cases}\label{eq57}
\end{align}
where $(K_{un},c_{un})$ denotes a pair of the lifted state gain and the constant term of the unconstrained optimal solution of $P_{\hat{z}_\infty}$, $(K_{\ell},c_{\ell})$ denotes a pair of the lifted state gain and the constant term of the optimal solution of $P_{\hat{z}_\infty}$ when the Lyapunov constraint in (\ref{eq50h}) is active at the optimum.
\end{lem}

\begin{pf}
Since the optimal control problem in (\ref{eq50}) is a quadratic program, the optimizer function is described as a piecewise affine function of the lifted state $\hat{z}_\infty$ \citep*{8,9}. Based on this fact, we derive the optimal solution for each case as a function of $\hat{z}_\infty$.

\quad\\
Case 1: We can reformulate the objective function in (\ref{eq50a}) of $P_{\hat{z}_\infty}$ as a quadratic function of $U=[u_0^\top, \cdots, u_{N-1}^\top ]^\top$:
\begin{align}
J_{\hat{z}_\infty}(U)=\frac{1}{2}U^\top H_\infty U +f_\infty^\top U \label{eq58}
\end{align}

When all the constraints are inactive at the optimum of $P_{\hat{z}_\infty}$, the following unconstrained optimal solution $U^*_{un}$ derived from the first-order optimality condition of $J_{\hat{z}_\infty}(U)$ is feasible and optimal:
\begin{align}
U^*_{un}=-H_\infty ^{-1} f_\infty \label{eq59}
\end{align}
where
\begin{align}
&H_\infty :=\Psi^\top Q_\textbf{z} \Psi+Q_\textbf{u} \nonumber\\
&f_\infty :=\Psi^\top Q_\textbf{z}(\Phi \hat{z}_\infty+\Psi_d\hat{\mathbf{d}}_\infty -\bar{\mathbf{z}}_\infty)-Q_\textbf{u} \bar{\mathbf{u}}_\infty \nonumber\\
&\Psi:=\begin{bmatrix} B & 0 & \cdots & 0 \\ AB & B & \cdots & 0 \\ \vdots & \vdots & \ddots & \vdots \\ A^{\textit{N}-1}B & A^{\textit{N}-2}B & \cdots & B \end{bmatrix} \nonumber\\
&\Psi_d:=\begin{bmatrix} B_d & 0 & \cdots & 0 \\ AB_d & B_d & \cdots & 0 \\ \vdots & \vdots & \ddots & \vdots \\ A^{\textit{N}-1}B_d & A^{\textit{N}-2}B_d & \cdots & B_d \end{bmatrix} \nonumber\\
&\Phi:=\begin{bmatrix} A \\ A^2 \\ \vdots \\ A^{N} \end{bmatrix},\; 
\hat{\mathbf{d}}_\infty:=\begin{bmatrix} \hat{d}_\infty \\ \hat{d}_\infty \\ \vdots \\ \hat{d}_\infty \end{bmatrix},\; 
\bar{\mathbf{z}}_\infty:=\begin{bmatrix} \bar{z}_\infty \\ \bar{z}_\infty \\ \vdots \\ \bar{z}_\infty \end{bmatrix},\;
\bar{\mathbf{u}}_\infty:=\begin{bmatrix} \bar{u}_\infty \\ \bar{u}_\infty \\ \vdots \\ \bar{u}_\infty \end{bmatrix} \nonumber
\end{align}
\begin{align}
& Q_\textbf{z}:=diag\{ Q_z,Q_z,\cdots,Q_z \}\nonumber\\
& Q_\textbf{u}:=diag\{ Q_u,Q_u,\cdots,Q_u \}\nonumber
\end{align}
where $diag\{ Q_1,Q_2,\cdots,Q_n \}$ represents a block diagonal matrix that has $Q_1,Q_2,\cdots,Q_n$ as the main diagonal blocks. Then, the control law $u_\infty$ can be described as the first component of $U^*_{un}$:
\begin{align}
&u_\infty=C_1 U^*_{un} \label{eq60}\\
&C_1:=\begin{bmatrix} I_{n_u} & 0 & \cdots & 0 \end{bmatrix}. \nonumber
\end{align}
By substituting (\ref{eq59}) into (\ref{eq60}) and rearranging, we obtain
\begin{align}
u_\infty=K_{un} \hat{z}_\infty +c_{un} \label{eq61}
\end{align}
where
\begin{align}
&K_{un}:=-C_1(\Psi^\top Q_\textbf{z} \Psi+Q_\textbf{u})^{-1}\Psi^\top Q_\textbf{z}\Phi\nonumber\\
&c_{un}:=C_1(\Psi^\top Q_\textbf{z} \Psi+Q_\textbf{u})^{-1}(\Psi^\top Q_\textbf{z}(\bar{\textbf{z}}_\infty-\Psi_d \hat{\textbf{d}}_\infty)+ Q_\textbf{u} \bar{\mathbf{u}}_\infty)\nonumber
\end{align}

\quad\\
Case 2: When the Lyapunov constraint in (\ref{eq50h}) is active at the optimum of $P_{\hat{z}_\infty}$, the control law $u_\infty$ can be described as
\begin{align}
u_\infty=K_{\ell}\hat{z}_\infty+c_{\ell}\label{eq62}
\end{align}
where $K_{\ell}$ and $c_{\ell}$ are derived in \textit{Appendix A}.

From (\ref{eq61}) and (\ref{eq62}), we can see that (\ref{eq56}) holds.\qed
\end{pf}

In \textbf{Lemma~\ref{lem3}}, we derived the relation between $\hat{z}_\infty$ and $\hat{u}_\infty$ by analyzing $P_{\hat{z}_\infty}$. Now, we derive the relation between the target lifted state $\bar{z}_\infty$ and target input $\bar{u}_\infty$ at steady state in the following lemma by analyzing $P(\bar{z}_\infty,\hat{d}_\infty,\bar{z}_\infty,\bar{u}_\infty)$ where the estimated lifted state $\hat{z}_\infty$, which is set as the initial lifted state $z_0$ in $P_{\hat{z}_\infty}$, is replaced with the target lifted state $\bar{z}_\infty$.

\begin{lem}\label{lem4}
Let $P_{\bar{z}_\infty}$ represent the optimal control problem $P(\bar{z}_\infty,\hat{d}_\infty,\bar{z}_\infty,\bar{u}_\infty)$. We assume that the process constraints in (\ref{eq50e}) and (\ref{eq50f}) and the first Lyapunov constraint in (\ref{eq50g}) are inactive at the optimum of $P_{\bar{z}_\infty}$. Then, $\bar{u}_\infty$ becomes the optimal control law of $P_{\bar{z}_\infty}$, and the following equation holds:
\begin{align}
\bar{u}_\infty=K_{mpc}\bar{z}_\infty +c_{mpc}.\label{eq63}
\end{align}
\end{lem}

\begin{pf}
Since the initial lifted state of $P_{\bar{z}_\infty}$ is set as $z_0=\bar{z}_\infty$, $\bar{\textbf{u}}_\infty:=[\bar{u}_\infty^\top,\cdots,\bar{u}_\infty^\top]^\top$ makes the future lifted states remain at $\bar{z}_\infty$ (i.e., $z_1,\cdots,z_N=\bar{z}_\infty$) from (\ref{eq54}). Then, since the target lifted state and target input of $P_{\bar{z}_\infty}$ are set as $\bar{z}_\infty$ and $\bar{u}_\infty$, respectively, eventually the value of the objective function in (\ref{eq50a}) of $P_{\bar{z}_\infty}$ becomes 0 with $\bar{\textbf{u}}_\infty$. Therefore, $\bar{u}_\infty$ is identical to the unconstrained optimal control law of $P_{\bar{z}_\infty}$ that yields the minimum objective value in the absence of the constraints.

Since all the parameters in $P_{\hat{z}_\infty}$ and $P_{\bar{z}_\infty}$ are identical except the initial lifted state values ($z_0=\hat{z}_\infty$ in $P_{\hat{z}_\infty}$ and $z_0=\bar{z}_\infty$ in $P_{\bar{z}_\infty}$, respectively), the unconstrained control law of $P_{\bar{z}_\infty}$ can be described with $\bar{z}_\infty$ using the same $K_{un}$ and $c_{un}$ of $P_{\hat{z}_\infty}$ in (\ref{eq61}):
\begin{align}
\bar{u}_\infty=K_{un}\bar{z}_\infty +c_{un}.\label{eq64}
\end{align}

Now, we analyze the behavior of the Lyapunov constraint in (\ref{eq50h}) in $P_{\bar{z}_{\infty}}$ when $U=\bar{\textbf{u}}_\infty$. For $P_{\bar{z}_\infty}$, the stabilizing control law in (\ref{eq44}) yields
\begin{align}
h(\bar{z}_\infty,\hat{d}_\infty,\bar{z}_\infty)=(\bar{N}-K_z)\bar{z}_\infty-K_d\hat{d}_\infty.\label{eq65}
\end{align}
Substituting $z_0=\bar{z}_\infty$ and (\ref{eq65}) into (\ref{eq50d}) yields
\begin{align}
z_1^h=(A-BK_z+B\bar{N})\bar{z}_\infty+(B_d-BK_d)\hat{d}_\infty.\label{eq66}
\end{align}
We can reformulate (\ref{eq66}) as
\begin{align}
z_1^h-\bar{z}_\infty+(I-A+BK_z)\bar{z}_\infty=B\bar{N}\bar{z}_\infty+(B_d-BK_d)\hat{d}_\infty.\label{eq67}
\end{align}
Then, multiplying $(I-A+BK_z)^{-1}$ on both the sides of (\ref{eq67}) and substituting $\bar{N}$ and $K_d$ in (\ref{eq45}) and (\ref{eq46}) into (\ref{eq67}) yields
\begin{align}
(I-A+BK_z)^{-1}(z_1^h-\bar{z}_\infty)=0.\label{eq68}
\end{align}
Since $I-A+BK_z$ is nonsingular, (\ref{eq68}) holds when (\ref{eq69}) is true.
\begin{align}
z_1^h=\bar{z}_\infty.\label{eq69}
\end{align}
Since the lifted state evolved from $\bar{z}_\infty$ with $\bar{u}_\infty$ under the influence of $\hat{d}_\infty$ becomes $\bar{z}_\infty$ as in (\ref{eq54}), (\ref{eq70}) holds when $z_1$ is derived from (\ref{eq50c}) with $u_0=\bar{u}_\infty$:
\begin{align}
z_1=\bar{z}_\infty.\label{eq70}
\end{align}

Now, by substituting $z_1^h$ and $z_1$ in (\ref{eq69}) and (\ref{eq70}) into (\ref{eq50h}), both sides of (\ref{eq50h}) become 0 and the equality is satisfied. This means that the Lyapunov constraint in (\ref{eq50h}) is active when $U=\bar{\textbf{u}}_\infty$. Therefore, $\bar{\textbf{u}}_\infty$ is the unconstrained optimal solution of $P_{\bar{z}_\infty}$ as shown in (\ref{eq64}) as well as the constrained optimal solution of $P_{\bar{z}_\infty}$ when the Lyapunov constraint in (\ref{eq50h}) is active. Since $P_{\bar{z}_\infty}$ has the same formulation as $P_{\hat{z}_\infty}$ with a different initial lifted state $\bar{z}_\infty$ as described above, we can describe $\bar{u}_\infty$ with the same $K_{\ell}$ and $c_{\ell}$ in (\ref{eq62}):
\begin{align}
\bar{u}_\infty=K_{\ell}\bar{z}_\infty +c_{\ell}.\label{eq71}
\end{align}

From (\ref{eq64}) and (\ref{eq71}), we can see $\bar{u}_\infty$ is the optimal control law of $P_{\bar{z}_\infty}$ in both cases $\mathcal{I}_A=\emptyset$ or $\mathcal{I}_A=\{ i_\ell \}$, and thus, (\ref{eq63}) holds.\qed
\end{pf}
 
Based on \textbf{Lemma~\ref{lem3}} and \textbf{Lemma~\ref{lem4}}, we can derive the relation between $u_\infty-\bar{u}_\infty$ and $\hat{z}_\infty-\bar{z}_\infty$ by subtracting (\ref{eq63}) from (\ref{eq56}):
\begin{align}
u_\infty-\bar{u}_\infty = K_{mpc} (\hat{z}_\infty-\bar{z}_\infty).\label{eq72}
\end{align}

Now, the condition and proof for zero-offset at steady state of the proposed offset-free KLMPC system are given in the following theorem.

\begin{thm}\label{thm5}
If $L_z$ and $L_d$ satisfy the null space condition in (\ref{eq73}), then the proposed offset-free KLMPC system yields zero-offset at steady state.
\begin{align}
&\mathcal{N}(L_d) \subseteq \mathcal{N}(H(I-C(I-A-BK_{mpc})^{-1}L_z)\label{eq73}
\end{align}
where $\mathcal{N}$ denotes null space.
\end{thm}

\begin{pf}
First, we derive the relation between $\hat{z}_\infty$ and $\bar{z}_\infty$. Subtracting (\ref{eq54}) from (\ref{eq51}) and rearranging yields
\begin{align}
(I-A)(\hat{z}_\infty-\bar{z}_\infty)=B(u_\infty-\bar{u}_\infty)-L_ze_{y,\infty}.\label{eq74}
\end{align}
Then, substituting (\ref{eq72}) into (\ref{eq74}) and rearranging yields
\begin{align}
\hat{z}_\infty-\bar{z}_\infty=-(I-A-BK_{mpc})^{-1}L_ze_{y,\infty}.\label{eq75}
\end{align}

Now, we examine the offset of the measured controlled variable $Hy_{p,\infty}$ from the set-point $\bar{y}_c$ at steady state:
\begin{align}
e_{y_c,\infty}=Hy_{p,\infty}-\bar{y}_c.\label{eq76}
\end{align}
By substituting (\ref{eq55}) into (\ref{eq76}), we obtain
\begin{align}
e_{y_c,\infty}=H(y_{p,\infty}-C\hat{z}_\infty-C_d\hat{d}_\infty+C(\hat{z}_\infty-\bar{z}_\infty)).\label{eq77}
\end{align}
Then, substituting (\ref{eq53}) and (\ref{eq75}) into (\ref{eq77}) yields
\begin{align}
e_{y_c,\infty}=H[I-C(I-A-BK_{mpc})^{-1}L_z]e_{y,\infty}.\label{eq78}
\end{align}
Combining (\ref{eq52}) and (\ref{eq78}) and rearranging yields
\begin{align}
\begin{bmatrix} 0 \\ I \end{bmatrix}e_{y_c,\infty} = \begin{bmatrix} L_d \\ H(I-C(I-A-BK_{mpc})^{-1}L_z) \end{bmatrix} e_{y,\infty}.\label{eq79}
\end{align}

Then, from the null space condition in (\ref{eq73}), $e_{y,\infty}$ that satisfies $L_d e_{y,\infty}=0$ also satisfies $H(I-C(I-A-BK_{mpc})^{-1}L_z)e_{y,\infty}=0$. Eventually, the second component of the right hand side of (\ref{eq79}) becomes 0, and thus the following equation holds:
\begin{align}
e_{y_c,\infty}=0.\label{eq80}
\end{align}
From (\ref{eq80}), we can see that the offset-free KLMPC system yields zero-offset at steady state.\qed
\end{pf}

\section{Numerical example}\label{sec3}

In this section, a numerical simulation is performed to demonstrate the effectiveness of the proposed offset-free KLMPC framework in handling the plant-model mismatch inherent in the nominal KLMPC. A continuous stirred-tank reactor (CSTR) in \cite*{4} where a first-order reaction takes place is considered for the closed-loop simulation. The dynamics of the CSTR is as follows:
\begin{subequations}\label{eq81}
\begin{align}
&\frac{dc}{dt}=\dfrac{F_0 (c_0-c)}{\pi r^2 h}-k_0e^{-\frac{E}{RT}}c\label{eq81a} \\
&\dfrac{dT}{dt}=\dfrac{F_0 (T_0 -T)}{\pi r^2 h}-\dfrac{\Delta H}{\rho C_p}k_0e^{-\frac{E}{RT}}c +\dfrac{2U}{r\rho C_p}(T_c -T) \label{eq81b}\\
&\dfrac{dh}{dt}=\dfrac{F_0-F}{\pi r^2}\label{eq81c}
\end{align}
\end{subequations}
where the parameters are given in Table~\ref{table1}.

\begin{table}[h]
\begin{center} 
\caption{Parameters for the CSTR considered in the simulation.}
\begin{tabular}{l|l|l}
\hline
Parameter & Value & Unit \\ \hline
Inlet flowrate, $F_0$ & 0.1 & $\mathrm{m^3 /min}$ \\
Inlet temperature, $T_0$ & 350 & $\mathrm{K}$\\
Inlet concentration, $c_0$ & 1 & $\mathrm{kmol/m^3}$\\
Frequency factor, $k_0$ & 7.2$\times 10^{10}$ & $\mathrm{min^{-1}}$\\
Specific heat, $C_p$ & 0.239 & $\mathrm{kJ/kg\cdot K}$\\
Activation energy, $E$ & 7.275$\times 10^4$ & $\mathrm{kJ/kmol}$\\
Gas constant, $R$ & 8.314 & $\mathrm{kJ/kmol\cdot K}$\\
Heat coefficient, $U$ & 54.94 & $\mathrm{kJ/min\cdot m^2 \cdot K}$\\
Radius of reactor, $r$ & 0.219 & $\mathrm{m}$\\
Density, $\rho$ & 1000 & $\mathrm{kg/m^3}$\\
Heat of reaction, $\Delta H$ & -5$\times 10^4$ & $\mathrm{kJ/kmol}$\\ \hline 
\end{tabular}\label{table1}
\end{center}
\end{table}

The control objective of the control system is to track the set-point values of the outlet concentration ($c$) and the reactor temperature ($T$) by manipulating the jacket temperature ($T_c$) and the outlet flow rate ($F$) of the reactor.

The closed-loop simulation is performed using the MATLAB \circledR R2019b with Intel \circledR Core TM i7-10510U CPU @ 1.80 GHz and 16GB RAM.

\subsection{Data-driven model identification through EDMD}\label{sec3_1}

To demonstrate the application of the proposed offset-free KLMPC framework, the nonlinear model of the CSTR in (\ref{eq81}) is utilized as a virtual plant, but the detailed dynamics of the CSTR is assumed to be not available.

To identify the linear state space model using EDMD in Section~\ref{sec1_1}, we collected input ($T_c$, $F$) and state ($c$, $T$, $h$) data from the virtual CSTR. Specifically, 1000 operation trajectories are generated by solving the ODEs in (\ref{eq81}) for 500 min for each operation with a 1 min sampling time interval. The initial condition for each operation is randomly assigned around the following desired steady-state:
\begin{align}\label{eq82}
x_s=[0.878, 324.5, 0.659].
\end{align}
The input trajectory for each operation is also randomly generated from the range determined by the following input constraints:
\begin{align}
290\leq T_c \leq 315,\; 0.04 \leq F \leq 0.16 \nonumber.
\end{align}

An observable function library $\psi(x)$ in (\ref{eq8}) is constructed as follows:
\begin{align}\label{eq83}
\psi(x)=[& x_1 , x_2 , x_3 , x_1^2 , x_2^2 , x_1x_2, \nonumber\\
&x_1e^{-1/x_2}, (x-x_s)^\top P (x-x_s) ]^\top.
\end{align}
Based on the discussion in Section~\ref{sec2_1}, the state itself and the Lyapunov function with the equilibrium point $x_s$ are included as observable functions. Utilizing these observables, the Lyapunov constraints can be transformed into linear forms as in (\ref{eq50g}) and (\ref{eq50h}).

\begin{figure}[h!]
\begin{center}
{\includegraphics[width=9cm]{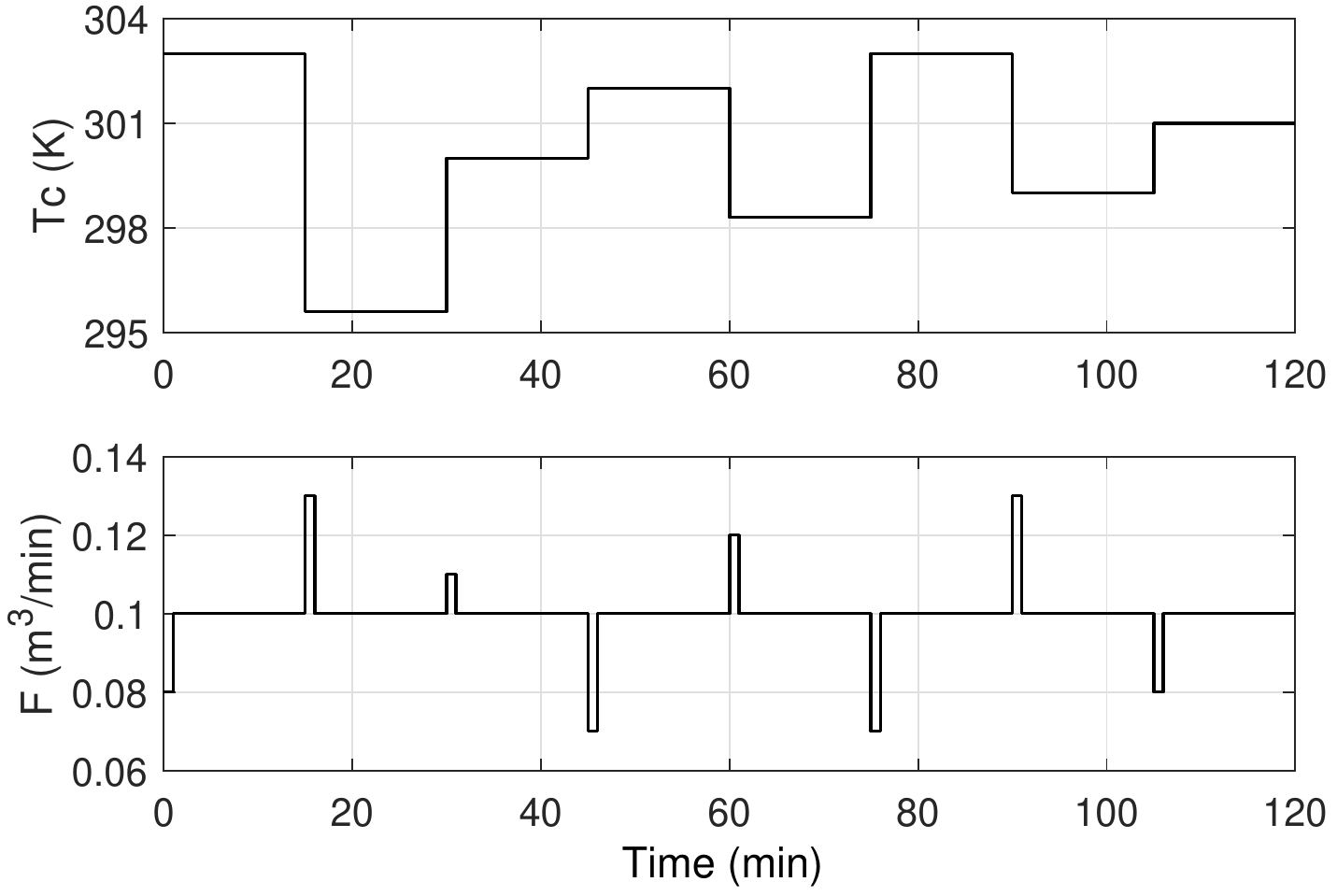}}
\end{center}
\caption{Input trajectories applied for validation of the linear dynamical model obtained via EDMD.}
\label{fig1}
\end{figure}

Then, a linear state-space model in the lifted space in (\ref{eq13}) is identified using the 500,000 sets of collected $((x_j,u_j),(x_j^+,u_j^+))$ data and the constructed observables. To validate the prediction power of the lifted state-space model, we considered a series of step inputs in $T_c$ and pulse inputs in $F$, with a duration of 15 min, as described in Fig.~\ref{fig1}.

\begin{figure}[h!]
\begin{center}
{\includegraphics[width=9cm]{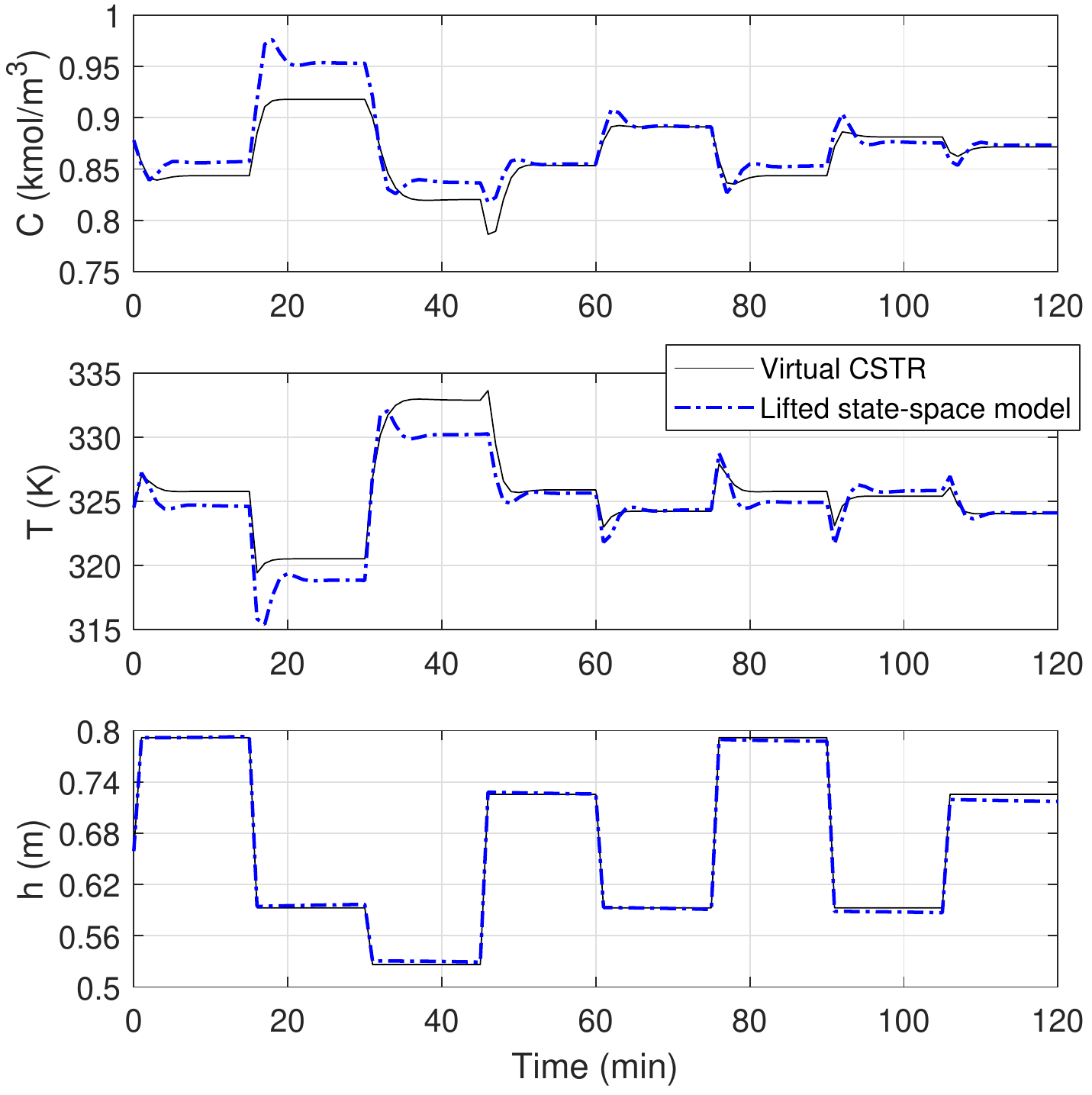}}
\end{center}
\caption{Output trajectories from the virtual CSTR and the lifted state-space model induced by the input trajectories in Fig.~\ref{fig1}.}
\label{fig2}
\end{figure}

Fig.~\ref{fig2} shows the output trajectories from the virtual CSTR in (\ref{eq81}) and the lifted state-space model in (\ref{eq13}) induced by the input trajectories in Fig.~\ref{fig1}. While the lifted state-space model seems to predict the behavior of the virtual CSTR for the most part utilizing the enriched observable space with nonlinear functions as in (\ref{eq83}), in Fig.~\ref{fig2}, prediction error still exists over operating time intervals [15 30] and [30 45] mins due to the inherent plant-model mismatch caused by the finite-dimensional approximation of the infinite-dimensional Koopman operator via EDMD. 

To quantitatively examine the prediction performance of the lifted state-space model, we derive a normalized root mean squared error (NRMSE) value for each output variable as follows:
\begin{subequations}
\begin{align}
&NRMSE:=\frac{RMSE}{y_{p,max}-y_{p,min}}\label{eq84a} \\
&RMSE:=\left( \frac{\sum^{n_D}_{j=1} (\hat{y}_{m,j}-y_{p,j})^2}{n_D} \right)^{1/2}\label{eq84b}
\end{align}
\end{subequations}
where $y_{p}$ and $\hat{y}_m$ denote the output from the virtual CSTR and the lifted state-space model, respectively. The resultant NRMSE value for each output is obtained as
\begin{align}
NRMSE_y=[0.1319, 0.0969, 0.0142]. \nonumber
\end{align}
The NRMSE value of $h$ is considerably smaller than those values of $C$ and $T$, and we can also see the prediction accuracy of $h$ is much higher than other variables in Fig.~\ref{fig2}. This can be attributed to the relatively simple and linear dynamics of $h$ in (\ref{eq81c}) that allows EDMD to derive an accurate linear predictor for $h$.

\subsection{Closed-loop simulation results}\label{sec3_2}

Based on the lifted state-space model derived from data-driven EDMD, we design an offset-free KLMPC system for reference tracking of the CSTR.

As previously discussed in Section~\ref{sec2}, a disturbance model is augmented with the Koopman-based lifted state space model with proper $B_d$ and $C_d$ that satisfy the condition in \textbf{Proposition~\ref{prp1}} to handle the plant-model mismatch effect. Subsequently, a disturbance estimator in (\ref{eq29}), a target problem in (\ref{eq33}), and an optimal control problem in (\ref{eq50}) are designed based on the offset-free KLMPC formulation in Section~{\ref{sec2}}. The prediction horizon of the model predictive controller is set as $N=10$. Additionally, the following process constraints for output variables are applied to the optimal control problem:
\begin{align}
&0.81\leq c\leq 0.92,\; 320\leq T\leq 330,\; 0.4 \leq h\leq 1.2. \nonumber
\end{align}

\begin{figure}[h!]
\begin{center}
{\includegraphics[width=9cm]{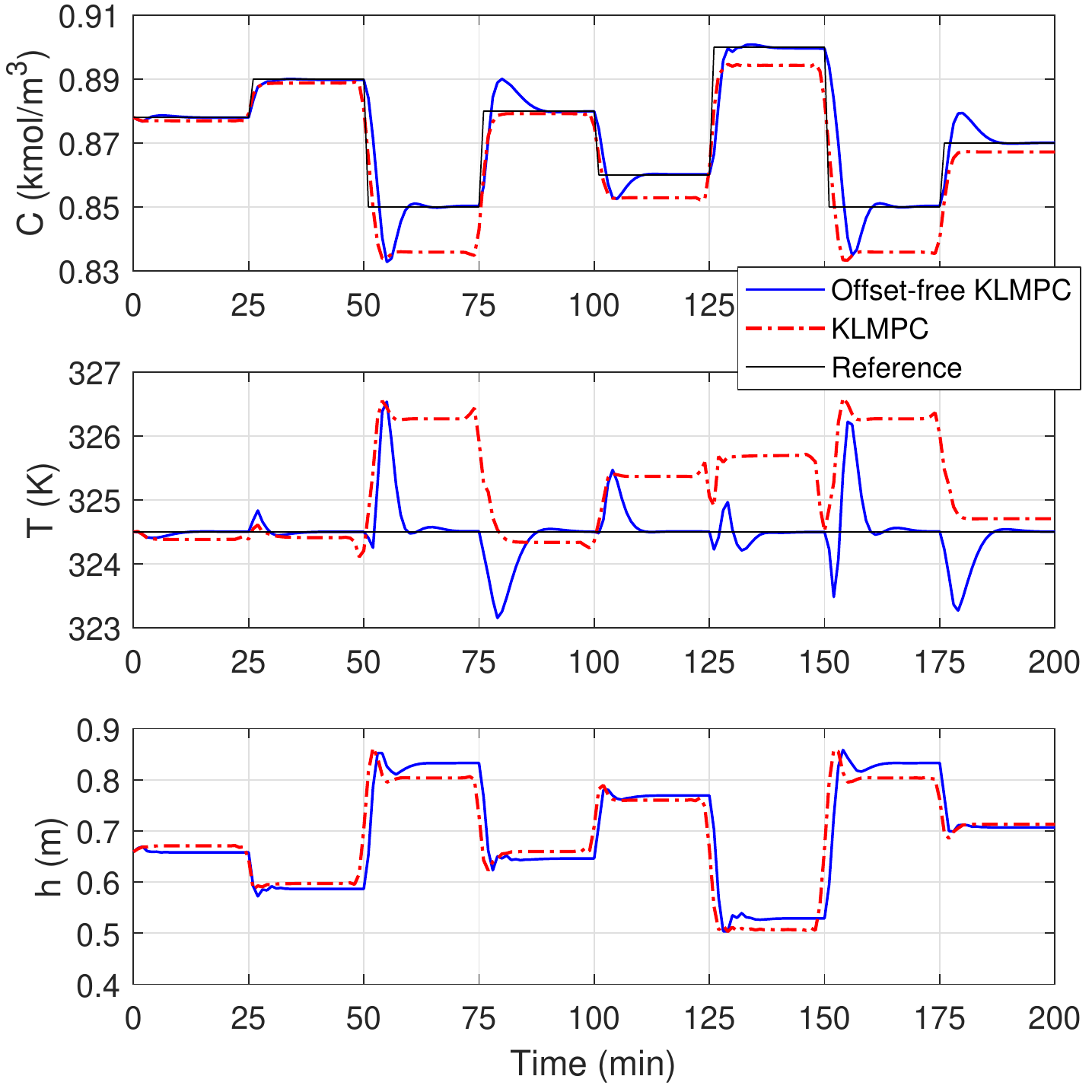}}
\end{center}
\caption{Reference tracking results from the proposed offset-free KLMPC and the nominal KLMPC systems.}
\label{fig3}
\end{figure}

Fig.~\ref{fig3} presents the closed-loop trajectories of the output variables induced by the proposed offset-free KLMPC and the nominal KLMPC schemes. The set-point value for $c$ is changed for every 25 min over [0.85, 0.90] while the set-point value for $T$ is fixed at 324.5 K. The measured $h$ value is applied to the disturbance estimator, but it is not treated as a controlled variable. We can see that $c$ and $T$ trajectories induced by the nominal KLMPC scheme show considerable offsets from the set-point values due to the plant-model mismatch between the linear model derived by EDMD in the previous section and the CSTR process in (\ref{eq81}). On the other hand, $c$ and $T$ trajectories induced by the proposed offset-free KLMPC scheme show zero steady-state offset.

\begin{figure}[h!]
\begin{center}
{\includegraphics[width=9cm]{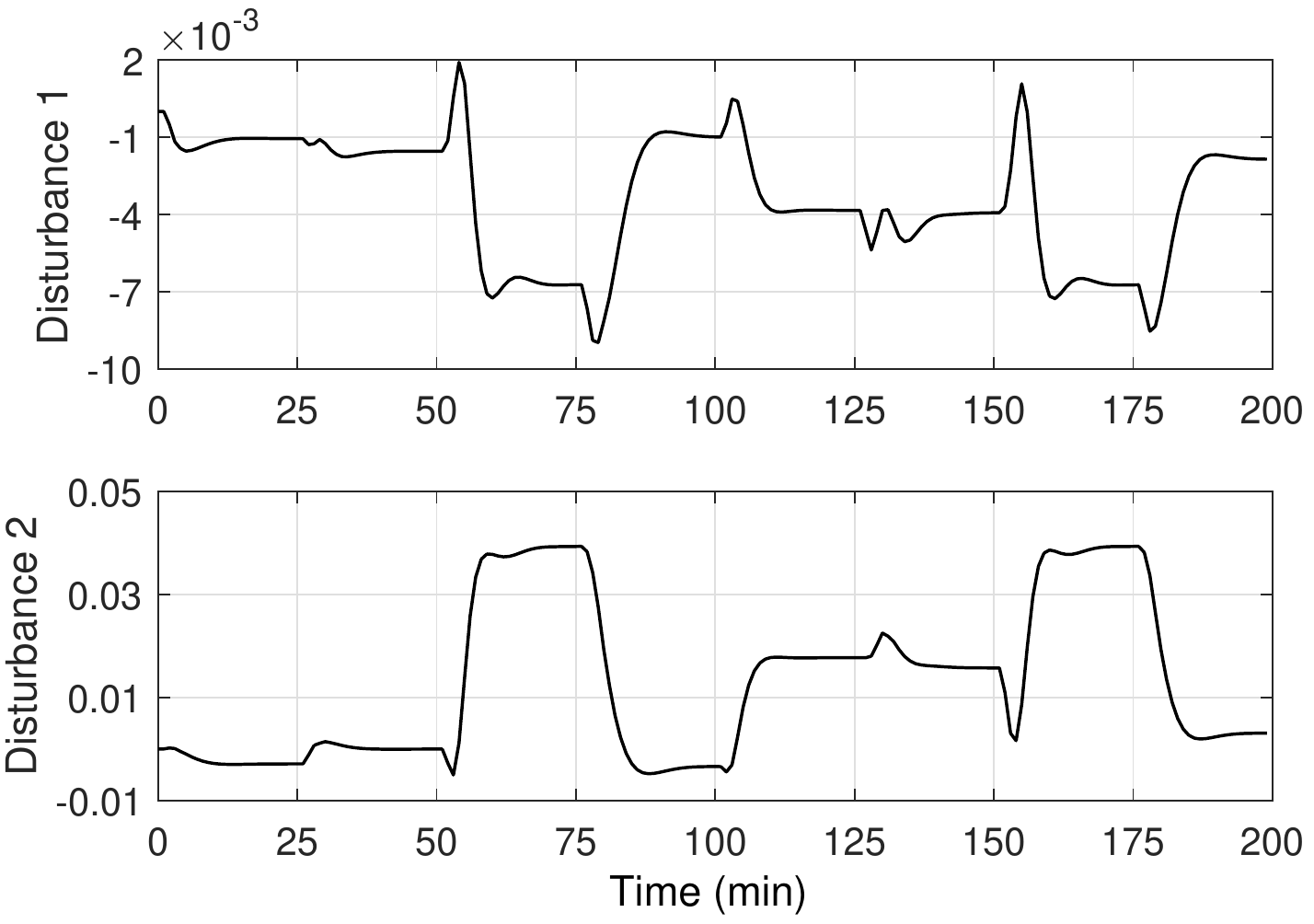}}
\end{center}
\caption{Estimated disturbance trajectories of the proposed offset-free KLMPC scheme.}
\label{fig4}
\end{figure}

This is because the proposed offset-free KLMPC framework can effectively compensate for the influence of plant-model mismatch by properly estimating the disturbance signals as shown in Fig.~\ref{fig4} and applying them to the control system. When the set-point changes, it takes some time to compute disturbance values via the estimator as shown in Fig.~\ref{fig4}, e.g., around 50, 75 and 150 mins. Eventually, the trajectories of controlled variables $C$ and $T$ in Fig.~\ref{fig3} show considerable deviations from the set-point values near those sampling instants. Nonetheless, as the control system reaches a steady state for each set-point, the disturbance values derived from the properly designed stable estimator converge to proper values as shown in Fig.~\ref{fig4}. By applying these disturbance values to the control system, the proposed offset-free KLMPC scheme can derive an optimal input to drive the CSTR system to a steady state where the zero steady-state offset is satisfied in the presence of plant-model mismatch on the controlled variables.

\section*{Conclusions}
In this work, we proposed the offset-free KLMPC framework to address plant-model mismatch in the KLMPC scheme which is inherently occurred in data-driven model identification through EDMD caused by a finite-dimensional approximation to the infinite-dimensional Koopman operator.

Specifically, a disturbance model is augmented with the lifted state-space model derived from EDMD to consider the plant-model mismatch. Then, the offset-free KLMPC framework that consists of an estimator, a target problem, and an optimal control problem is developed. The estimator derives lifted state and disturbance estimates from the measured prediction error, and the target problem derives proper targets for lifted state and input considering the effect of disturbance. Then, the optimal control problem with Lyapunov constraints obtains an optimal input to track the set-point while compensating for the plant-model mismatch, and ensuring feasibility and stability of the control system.

Unlike the nominal KLMPC scheme, since the equilibrium point of the optimal control problem (i.e., steady-state target) is continuously updated with the disturbance estimate in the proposed framework, the formulations for updating Lyapunov function and stabilizing control law in Lyapunov constraints are also developed. Additionally, a mathematical analysis for the zero steady-state offset condition of the proposed framework is performed.

The numerical simulation results showed the effectiveness of the proposed offset-free KLMPC framework in addressing the plant-model mismatch compared to the nominal KLMPC scheme. This is especially meaningful because the inaccuracy in prediction due to the plant-model mismatch can be exacerbated when a system state is lifted to the space of observables. In conclusion, the developed offset-free KLMPC framework is expected to provide a valuable direction for addressing the plant-model mismatch that is inherent in Koopman operator-based predictive control methods.
	
\section*{Acknowledgments}
	Financial support from the Artie McFerrin department of chemical engineering and the Texas A\&M Energy Institute are gratefully acknowledged.
	
\appendix

\section{Derivation of (\ref{eq62})}
We can reformulate $P_{\hat{z}_\infty}$ in a compact form by expressing the predicted lifted states in the objective function and inequality constraints explicitly with $U=[u_0^\top, \cdots, u_{N-1}^\top ]^\top$:
\begin{subequations}\label{eqA1}
\begin{align}
P_{\hat{z}_\infty}:\; \underset{\small{U}}{\mathrm{min}}&\quad \frac{1}{2}U^\top H_\infty U +f_\infty^\top U \label{eqA1a}\\
\mathrm{s.t.}&\quad G_\infty U\leq g_\infty+ S_\infty \hat{z}_\infty\label{eqA1b}
\end{align}
\end{subequations}

Then, we can derive the first-order Karush-Kuhn-Tucker (KKT) optimality conditions of $P_{\hat{z}_\infty}$ as
\begin{subequations}\label{eqA2}
\begin{align}
&H_\infty U^*+f_\infty +G_\infty^\top v^*=0 \label{eqA2a}\\
&v_i^*(G_{\infty,i} U^* -g_{\infty,i} -S_{\infty,i} \hat{z}_\infty)=0\label{eqA2b}\\
&v_i^*\geq 0\label{eqA2c}\\
&G_\infty U^*-g_\infty- S_\infty \hat{z}_\infty\leq 0\label{eqA2d}
\end{align}
\end{subequations}
where $v$ denotes the dual variable, and $i$ denotes the index of constraints.

By rearranging (\ref{eqA2a}), we can obtain
\begin{align}
U^*=-H_\infty^{-1}(G_\infty^\top v^*+f_\infty).\label{eqA3}
\end{align}
Then, substituting (\ref{eqA3}) into (\ref{eqA2b}) yields
\begin{align}
v_i^*(-G_{\infty,i}H^{-1}_\infty(G_\infty^\top v^* +f_\infty)-g_{\infty,i}-S_{\infty,i} \hat{z}_\infty)=0. \label{eqA4}
\end{align}

When the Lyapunov function constraint in (\ref{eq50h}) is active at the optimum, (\ref{eqA5}) holds by complementary slackness.
\begin{align}
-G_{\infty,i_\ell}H^{-1}_\infty(G_{\infty,i_\ell}^\top v^*_{i_\ell} +f_{\infty})-g_{\infty,i_\ell}-S_{\infty,i_\ell} \hat{z}_\infty=0\label{eqA5}
\end{align}
where
\begin{align}
&G_{\infty,i_\ell}:=F_vBC_1 \nonumber\\
&g_{\infty,i_\ell}:=F_vB(\bar{N}\bar{z}_\infty-K_d \hat{d}_\infty) \nonumber\\
&S_{\infty,i_\ell}:=-F_vBK_z \nonumber\\
&F_v:=D_v+2(\bar{z}_s-\bar{z}_\infty)^\top D_x^\top Q_v D_x \nonumber\\
&C_1:=\begin{bmatrix} I_{n_u}, 0, \cdots, 0 \end{bmatrix}. \nonumber
\end{align}
Rearranging (\ref{eqA5}) yields
\begin{align}\label{eqA6}
v^*_{i_\ell}=-&(G_{\infty,i_\ell}H^{-1}_\infty G_{\infty,i_\ell}^\top)^{-1}\\
&(G_{\infty,i_\ell}H^{-1}_\infty f_{\infty}+g_{\infty,i_\ell}+S_{\infty,i_\ell} \hat{z}_\infty)  \nonumber
\end{align}

Then, substituting (\ref{eqA6}) and $v^*_i=0\;\forall i\neq i_\ell$ into (\ref{eqA3}) yields
\begin{align}
U^*=-H_\infty^{-1}(G_{\infty,i_\ell}^\top v^*_{i_\ell}+f_\infty).\label{eqA7}
\end{align}
Finally, we obtain $u_\infty$ from (\ref{eqA7}):
\begin{align}
u_\infty=C_1 U^*=K_{i_\ell}\hat{z}_\infty+c_{i_\ell}\label{eqA8}
\end{align}
where
\begin{align}
&K_{i_\ell}:=C_1 (M_1\Psi^\top Q_\mathbf{z} \Phi-M_2S_{\infty,i_\ell}) \nonumber\\
&c_{i_\ell}:=C_1 (M_1(\Psi^\top Q_\mathbf{z}(\Psi_d \hat{\mathbf{d}}_\infty-\bar{\mathbf{z}}_\infty)-Q_\mathbf{u}\bar{\mathbf{u}}_\infty )-M_2g_{\infty,i_\ell}) \nonumber\\
&M_1:=H_\infty^{-1}G_{\infty,i_\ell}^\top(G_{\infty,i_\ell}H^{-1}_\infty G_{\infty,i_\ell}^\top)^{-1}G_{\infty,i_\ell}H^{-1}_\infty -H_\infty^{-1}\nonumber\\
&M_2:=-H_\infty^{-1}G_{\infty,i_\ell}^\top(G_{\infty,i_\ell}H^{-1}_\infty G_{\infty,i_\ell}^\top)^{-1}. \nonumber
\end{align}
	
\bibliographystyle{model2-names}

\bibliography{references}
	
\end{document}